\documentclass[%
 aip,
rsi,%
 amsmath,amssymb,
 reprint,%
]{revtex4-1}
\usepackage{color}
\usepackage{graphicx}% Include figure files
\usepackage{dcolumn}% Align table columns on decimal point
\usepackage{bm}% bold math
\usepackage{xcolor}
\usepackage{ulem}
\begin{document}

\title{ Analog Signal Processing Using Stochastic Magnets }% Force line breaks with \\

\author{Samiran Ganguly}
\email{sganguly@virginia.edu}
 \affiliation{Charles L. Brown Dept. of Electrical and Computer Engineering, University of Virginia, Charlottesville, VA 22904}%Lines break automatically or can be forced with \\
\author{Kerem Y. Camsari}%
 \email{kcamsari@purdue.edu}
\affiliation{School of Electrical and Computer Engineering, Purdue University, West Lafayette, IN 47907}
\author{Avik W. Ghosh}
\email{ag7rq@virginia.edu}
\affiliation{Charles L. Brown Dept. of Electrical and Computer Engineering, University of Virginia, Charlottesville, VA 22904}
\affiliation{Department of Physics, University of Virginia, Charlottesville, VA 22904}


\begin{abstract}
We present a low barrier magnet based compact hardware unit for analog stochastic neurons and demonstrate its use as a building-block for neuromorphic hardware. By coupling circular magnetic tunnel junctions (MTJs) with a CMOS based analog buffer, we show that these units can act as leaky-integrate-and fire (LIF) neurons, a model of biological neural networks particularly suited for temporal inferencing and pattern recognition. We demonstrate examples of temporal sequence learning, processing, and prediction tasks in real time, as a proof of concept demonstration of scalable and adaptive signal-processors. Efficient non von-Neumann hardware implementation of such processors can open up a pathway for integration of hardware based cognition in a wide variety of emerging systems such as IoT, industrial controls, bio- and photo-sensors, and Unmanned Autonomous Vehicles. % \textcolor{blue}{(AG: Good tagline - "UAVs at UVA" !!)}.
\end{abstract}

                             %% Classification Scheme.
\maketitle

%\begin{quotation}
%The ``lead paragraph'' is encapsulated with the \LaTeX\ 
%\verb+quotation+ environment and is formatted as a single paragraph before the first section heading. 
%(The \verb+quotation+ environment reverts to its usual meaning after the first sectioning command.) 
%Note that numbered references are allowed in the lead paragraph.
%%
%The lead paragraph will only be found in an article being prepared for the journal \textit{Chaos}.
%\end{quotation}

\section{\label{sec:intro}Introduction}

Temporal inferencing and learning form the next frontier in the discipline of Artificial Intelligence. Development of hardware that can implement these tasks {\it{in-situ}}  can revolutionize the rapidly emerging era of smart-sensors, self-driving automotives, Unmanned Autonomous Vehicles (UAVs), and the Internet of Things (IoTs) by opening a pathway for self-contained, energy-efficient, highly scalable, and secure machine intelligence. In this work we propose a hybrid unit consisting of a low-barrier magnet based tunnel-junction (MTJ) coupled with a conventional CMOS based analog buffer as a building block for neuromorphic hardware that is particularly suited for temporal learning based signal processing tasks. 

MTJs lie at the heart of the Spin Transfer Torque based Magnetic Random Access Memory (STT-MRAM), a rapidly emerging commercial non-volatile memory technology, and can be built with semiconductor fabrication facilities available today \cite{xie_multiscale_2017}. The dynamics of our proposed hardware unit has a one-to-one mapping with the physical behavior of biological neurons, in particular stochastic leaky-integrate-and fire (LIF) neurons. We further show how networks assembled from these building-blocks can successfully learn and reproduce a chaotic signal by building temporal generative models, and work as adaptive filters by inverse modeling a communication channel with non-linear distortions. 

The ultra-compact footprint of these building-blocks with built-in neuron like behavior enables the design of reconfigurable large scale analog neuromorphic hardware that is more energy-efficient and highly scalable compared to present day practice, where neural networks are emulated as software models on Boolean algebra based hardware typically in cloud, with its associated inefficiencies, as well as concerns with cybersecurity and high communication bandwidth consumption.

\section{Stochastic Neural Network Nodes using Magnetic Tunnel Junctions}

\begin{figure}
\includegraphics[width=3.3in]{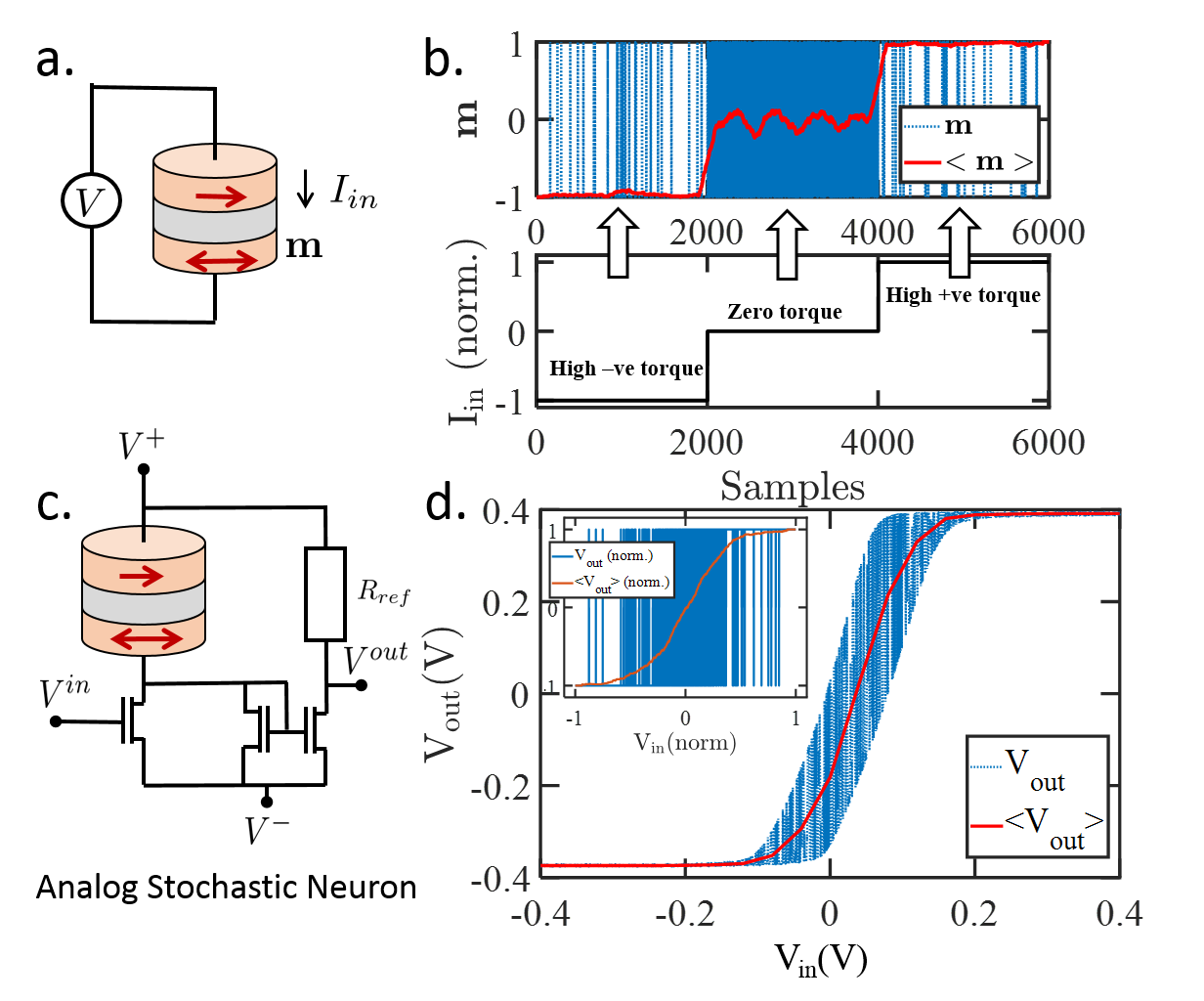}% Here is how to import EPS art
\caption{\label{fig:pbit} \textbf{Low barrier magnet MTJ based analog stochastic neuron:}  a. Circular low barrier magnet based MTJ that forms the heart of the ASN. b. The instantaneous and moving average response of the magnetization $\mathbf{m}$ of the MTJ free layer in presence of a high negative bias, zero bias, high positive bias. c. Analog stochastic neuron design. d. Instantaneous and time averaged output of the ASN under an input current sweep obtained from a SPICE model of the unit. Inset: instantaneous and averaged response of BSN.}
%\textcolor{blue}{(AG: Why not also show the Binary Stochastic Neuron here? Unit and Output, just to clarify? Also, what voltage $V_{DD}$ is here? Finally, what about voltage asymmetry of switching?)}}
\end{figure}

In magnets the state retention time is given by the Arrhenius relation \cite{kent_new_2015}:

\begin{equation}
\tau = \tau_0\exp(U/kT)
\label{eq:state-retention}
\end{equation}
For memory technology, the energy-barrier $U$ targeted for the MTJ free layer is at least $40kT$, to maintain a high state retention ($\tau \sim 10$ years) with typical $\tau_0\sim0.1-1\ ns$. The $U (= M_sH_k\Omega/2 )$ is determined by material properties such as saturation magnetization ($M_s$), anisotropy field strength ($H_k$), and geometrical volume ($\Omega$). However, in this work we use low energy-barrier magnets achieved by ultra-scaling \cite{debashis_experimental_2016} (fig.~\ref{fig:pbit}a) to enable fast dynamics in the reservoir and to leverage the built-in stochasticity provided by such magnets. In this case, the free layer magnetization $\mathbf{m}$ stays randomized between the two energy minima states when no current is provided. A large spin-torque from a large driving input current $I_{in}$ biases  $\mathbf{m}$ towards one of its minima directions, as shown in fig.~\ref{fig:pbit}b. We can utilize this controllable stochastic behavior to build noisy hardware neurons, both analog and digital versions.%\textcolor{red}{By using various electronic units in series with the MTJ, we can convert the magnetization to a suitable voltage that can replicate noisy digital bits (`p'-bits) discussed in the literature, as well as analog neuronal behavior particular suited for real time learning and prediction.}

\subsection{Analog and Binary Stochastic Neurons}

\subsubsection{Analog Stochastic Neuron (ASN)}

The core of the proposed hardware unit is a 1-MTJ 1-T in a pull-up, pull-down configuration as shown in fig.~\ref{fig:pbit}c. We then add a Wilson current mirror based analog buffer to this unit to generate a noisy analog output, while preventing any loading effects on the transfer characteristics of the unit from high fan-out at the output end. In this unit, the MTJ's intrinsic resistance ($RA = 100 \Omega -\mu m^2$, area $=\pi\times 50\times 50\ nm^2$, $TMR = 100 \%$ ), is chosen to be equal to the transistor resistance ($R = 12 k\Omega$) in the linear mode, while the $R_{ref}$ resistance is 0.2 times this value. When the input transistor is in saturation or cut-off mode, the voltage appearing at output (drawn from the drain on the transistor) is closer to $V_+$ or $V_-$ (where $ V_-=-V_+,V_+-V_- = V_{DD} = 0.8V$) respectively due to high bias voltage. In the linear mode of operation, the output of the unit goes through intermediate voltages as shown in fig.~\ref{fig:pbit}d. The response is inherently noisy due to the thermal noise's effect on the free layer magnetization as described before. Both the averaged signal and its upper and lower bounds show a $\tanh$ (or logistic function) like excitation. The overall response of the unit can be modeled by the following equation:

\begin{equation}
V_{out} = V_{DD}\tanh(\beta V_{in})/2 + \alpha(V_{in}) V_{rnd}
\label{eq:pbit_analog}
\end{equation}
where the parameters $\alpha, \beta$ depend on the particulars of device design. It can be be shown that a transistor's turn off and turn on depends on the deposition of a critical amount of switching charge on the gate terminal of the transistor \cite{datta_what_2014,ganguly_evaluating_2016}. This switching charge, in our design, is supplied by the net current from preceding neurons  flowing in to the resistive-capacitive metallic interconnects and then into the gate capacitor, where it automatically gets weighted, summed, and integrated over time ($Q_{in} = \int \displaystyle\frac{d(CV_{in})}{dt}dt = \int{I_{in}dt} $). The use of a low-barrier magnet in this structure inherently introduces volatility due to thermal noise, resulting in leakiness of the input current integration. Therefore, this unit behaves as a stochastic leaky-integrate-and-fire (sLIF) neuron. Additionally, adjustment of barrier height of the magnets ($U$) allows for tuning the dynamical rates of sLIF neural networks built from this unit.

\subsubsection{Binary Stochastic Neuron (BSN or p-bit)}

A binary stochastic neuron can be built from the same 1-MTJ 1-T unit. However, instead of a analog current mirror, we use a digital CMOS buffer (e.g. two cascaded NOT gates) as the output stage. This turns the output of the unit digital, i.e. $V_{out}$ is either $V_+$ or $V_-$, probability of which is dictated by a $\tanh$ law, unlike ASN whose response is continuous between $V_+$ and $V_-$ . The response of the BSN is shown in fig.~\ref{fig:pbit}d inset and given by: 
\begin{equation}
V_{out} = \mathrm{sgn}\Bigl(\tanh(\beta V_{in}) + \alpha(V_{in}) V_{rnd}\Bigr)V_{DD}/2
\label{eq:pbit_binary}
\end{equation}
This BSN design, presented elsewhere and called ``p-bit''  \cite{camsari_implementing_2017}, has been used in a variety of  optimization problems (see refs. \cite{sutton_intrinsic_2017,camsari_stochastic_2017}). These two related but distinct units, ASN and BSN, could form building blocks for a variety of neural networks, depending on the behavioral requirements. Controllability of behavioral noise through device design and electrical control make these units particularly useful in cases where stochasticity is an integral feature \cite{turchetti_stochastic_2004} in the computation (e.g. stochastic gradient descent\cite{lecun_efficient_2012}, Boltzmann machines\cite{coughlin_neural_1995}).

It should be noted that it is possible to build even more compact versions of these stochastic neurons by leveraging emerging phenomena and devices such as Giant Spin Hall Effect or Magneto-Electric based switching (e.g. see \cite{ganguly_evaluating_2016} for a few example designs). However, such devices are still in their research phase and their integration into larger circuits is still an open problem.

\section{\label{sec:rescomp}Computing Using Dynamical States: Computing Model and Hardware Implementation}

\subsection{\label{sec:res} Reservoir Computing: A Short Overview}

\begin{figure}
\includegraphics[width=2.5in]{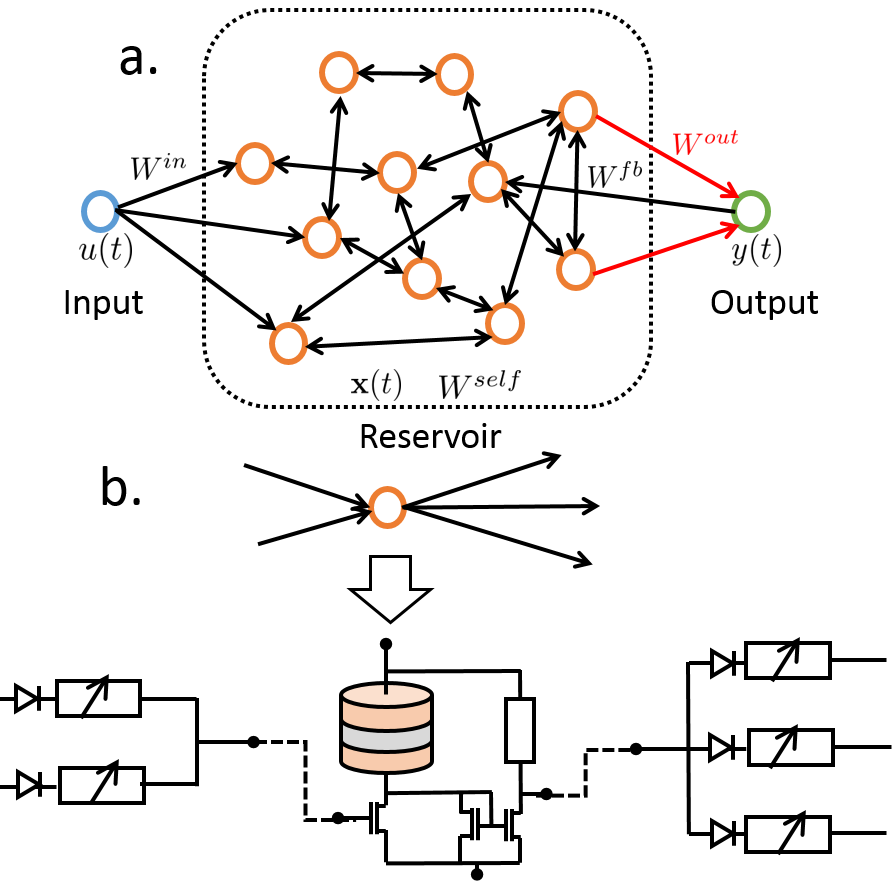}% Here is how to import EPS art
\caption{\label{fig:reservoir} \textbf{Hardware implementation of a reservoir computer:}  a.  The reservoir is a network of randomly connected sLIF neurons that processes an input stream and produces the output stream as a collective response. b. The reservoir node is built from an ASN, while the synaptic connections are made from controllable resistor networks.}
\end{figure}

Reservoir Computers (RC) are models of biological neural networks\cite{dominey_complex_1995,pascanu_neurodynamical_2011} that have been used for various signal processing tasks \cite{jaeger_harnessing_2004,triefenbach_phoneme_2010,jalalvand_real-time_2015,ganguly_hardware_2018}. In these networks, the computation is performed by a collection of randomly coupled  non-linear units with recurrent network topology (fig.~\ref{fig:reservoir}a). Such networks: (a) provide a huge expansion of the dynamical phase-space, increasing the distance between the signal-class centroids;  and (b) give rise to memory states in the network \cite{lukosevicius_practical_2012} allowing a signal to be temporally correlated, resulting in better signal classification.  The nodes of the network are leaky; therefore, the network memory is short term and fading - a feature critical to avoid overtraining. 

\textbf{\textit{RC Dynamics}}: Let $\mathbf{x}$ be the collective state vector of the reservoir, $\mathbf{u}$ be the input vector and $\mathbf{y}$ the output vector. Also let $W^{in}, W^{self}, W^{out}, W^{fb}$ be the matrices representing the synaptic connections between the input-reservoir, reservoir-reservoir, reservoir-output, and output-reservoir nodes respectively. The most general form of the RC dynamical equations is given by:

\begin{eqnarray}
\label{eq:dynamical1}
\frac{d\mathbf{x}}{dt} &=& -\eta \mathbf{x}+\kappa f_{NL}(W^{in} \mathbf{u}+W^{fb} \mathbf{y}+W^{self} \mathbf{x}) +\mathbf{\nu}\\
\mathbf{y} &=& W^{out} \mathbf{x}
\label{eq:dynamical2}
\end{eqnarray}
Here $\eta$ and $\kappa$ are system constants representing the leaking rate and the strength of the activation in the reservoir, $f_{NL}$ is a non-linear function, usually $\tanh$, and $\nu$ is the noise. This dynamical equation is equivalent to a model for a network of stochastic LIF neurons whose synaptic strengths are given by the various $W$ matrices.

\textbf{\textit{RC Learning}}: We use a weighted linear sum on $\mathbf{x}$ for time-series pattern learning and classification. The only synaptic weights adjusted during the training are the reservoir-output connection $W^{out}$ which involves minimizing the $\ell^2$ norm: $||W^{out}\mathbf{x}-\mathbf{y}||^2$ by finding the optimal $W^{out}$ (for example see Weiner-Hopf method\cite{lukosevicius_practical_2012})  .

\subsection{\label{sec:hardware} Reservoir Computer Hardware Implementation}

The RC dynamical equation (eq.~\ref{eq:dynamical1}) on a discretized temporal grid equispaced by $\delta t = 1$ in normalized time units (i.e., the magnet's dynamical time-scale ${\gamma H_k}/{(1+\alpha^2)}$) and $f_{NL} \equiv \tanh$, can be written as follows:

\begin{equation}
\label{eq:discretedyn1}
\mathbf{x}[t+1] = \underbrace{\kappa^\prime \tanh(\mathbf{z}[t+1])}_{\text{activation}} - \underbrace{(\eta^\prime-1) \mathbf{x}[t])}_{\text{leak}} -\underbrace{\mathbf{\nu^\prime}[t]}_{\text{noise}}
\end{equation}
where $^\prime$ denotes an additional factor of $\Delta t$, and $\mathbf{z}[t+1] = W^{in}\mathbf{u}[t+1] + W^{fb}\mathbf{y}[t] + W^{self}\mathbf{x}[t]$. Eq.~\ref{eq:discretedyn1} can be interpreted as describing a blackbox, whose output ($\mathbf{x}[t+1]$) is the sum of three terms: a) a transduction function given by a $\tanh$ type nonlinear activation, b) ``leaked'' past state $\mathbf{x}[t]$, where the leakiness arises from small state-retention times of ASN, c) noise inherent to the unit, both of which are naturally provided by the low-barrier magnets. %\textcolor{red}{captured} through a stochastic mechanism inherent in the node \textcolor{blue}{(AG: The leak is $\eta$ and the stochastic is $\nu$ - are we conflating the two here?)}. 
The ASN's electrical response directly corresponds with the behavior described by eq. \ref{eq:discretedyn1} and therefore it can be used to build compact reservoir computing nodes. For hardware based reservoir computing proposals using other material systems please see \cite{schrauwen_compact_2008,kulkarni_memristor-based_2012,larger_photonic_2012,van_der_sande_advances_2017,vandoorne_experimental_2014,torrejon_neuromorphic_2017,du_reservoir_2017,canaday_rapid_2018,opala_neuromorphic_2018,prychynenko_magnetic_2018}.

\subsection{Programmable Hardware Synapses}

A fully hardware based neural network necessitates that the physical interconnections or synaptic-weights be controllable in strength. In the presented hardware unit, the input signal is the net current flowing in the unit, while the output signal is the resulting voltage level. Therefore a resistor network can implement the synaptic weighing using Kirchoff's current law at each neuron's input, i.e. $I_{in} = \sum G_k V_{k}$, where interconnect conductances $G_k$'s are proportional to synaptic weights of the interconnection with other neurons (fig. \ref{fig:reservoir}b). Optionally, a small p-n junction diode can be introduced to ensure uni-directionality of current flow within the synaptic network with an added circuit cost of adjusting bias voltage ranges, since ASN and BSN require bipolar voltage range for operation. % \textcolor{blue}{(AG: What is the cost?)}. 

These programmable resistor networks could be implemented using a MOSFET in linear mode, whose channel resistance is controlled by the gate voltage.  Compact memristor cross-bar arrays \cite{jo_nanoscale_2010,kim_functional_2012} might be even better suited for this task since memristors are non-volatile and therefore more energy efficient than transistor networks.

%In most neural network training procedures, the central task involves $\ell^2$ norm minimization between the expected and generated outputs, which is equivalent to the task of finding the ground state of an Ising network \cite{ganguly_evaluating_2016}. The full implementation of  a purely hardware training circuit using such energy-minimizing circuits is left as a future work.

\section{\label{sec:tasks} Signal Processing Using Analog Stochastic Neuron (ASN) networks}

\subsection{Chaotic Time-Series Predictor}

\begin{figure}
\includegraphics[width=2.5in]{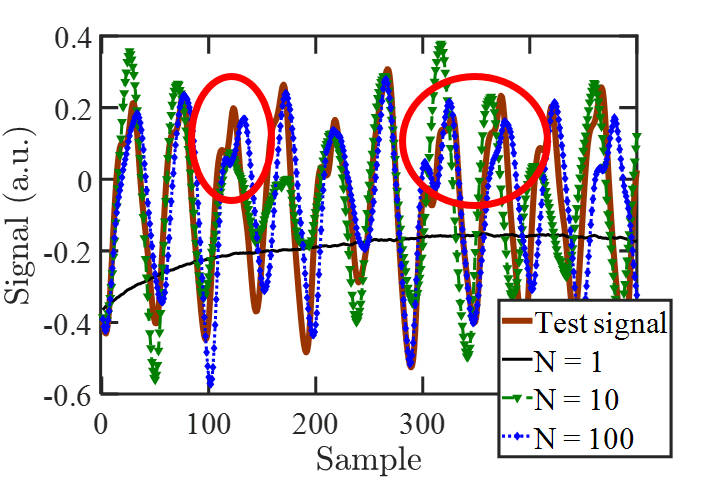}% Here is how to import EPS art
\caption{\label{fig:mgs} \textbf{Mackay-Glass (MG) system predictor using ASN network:} ASN network tested to generate chaotic time series. It can be seen from the lineshapes in the highlighted parts that larger networks generate a better fit with the MG.}
\end{figure}

The Mackey-Glass (MG) equation\cite{mackey_oscillation_1977} is a time series generator with periodic but subtly chaotic characteristics. The generating equation is given by:

\begin{equation}
\frac{dx}{dt}= \frac{\beta x(t-\tau)}{1+x(t-\tau)^n }-\gamma x
\label{eq:mgs}
\end{equation}
We train our ASN network to generate an MG system with a chaotic datastream for training, and then tested it on a test signal from the same generator. The ASN learns to reproduce the generator signal purely from its previously self-generated output. We found that for small number of nodes, the network fails to match the MG signal, but starts to generate better match for larger networks (see highlighted areas in fig. \ref{fig:mgs}). This happens because of the substantially richer dynamics and phase-space volume possible in a larger network. 

This task illustrates the possibility of creating temporal sequence-predictors and temporal auto-encoders using ASNs. Such temporal predictors and auto-encoders can find applications in temporal data modeling and reconstruction, and early-warning systems in bio-physical signal monitors by distinguishing out-of-the-norm patterns and beats, such as cardiac arrhythmia and seizures.

\subsection{Filtering Using Learning}

We now demonstrate a task at the heart of signal processing and digital wireless communication, i.e. signal filtering and channel equalization.  The task is to recover a bitstream after it passes through a medium or channel that introduces non-linear distortions, inter-symbol interference, and noise which cannot be fully compensated using a linear filter \cite{diniz_nonlinear_2013}. The principal idea behind our implementation of channel equalizer is to use an ASN network to reverse the effect of the channel, by learning the inverse of the underlying model of the channel's transfer function.

\begin{figure}
\includegraphics[width=2.5in]{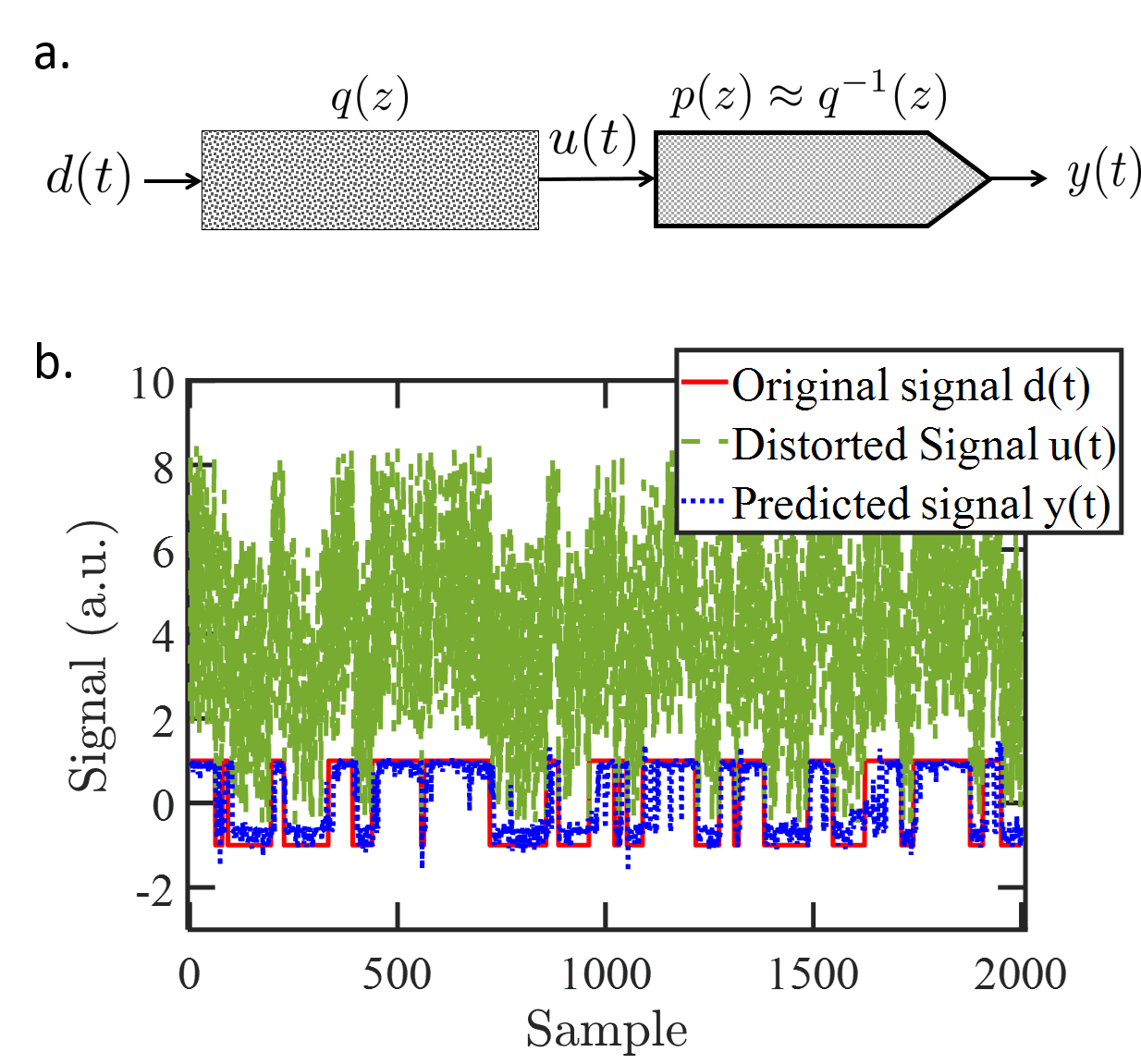}% Here is how to import EPS art
\caption{\label{fig:cheq} \textbf{A non-linear adaptive filter using ASN network:} a. Datastream $d(t)$ is passed through a noisy channel $q(z)$ that causes non-linear distortions to the datastream. b. A small ASN network can reproduce the original datastream $d(t)$ with high fidelity from the distorted data $u(t)$.}
\end{figure}

Let $d(t)$ be the original signal which goes through a channel (fig.\ref{fig:cheq}.a.) whose transfer function $p(z)$ produces $u(t)=p(d(t))$ and is given by:

\begin{equation}
u(t)=\sum_m \sum_n A_n  [\sum_k B_k[d(t-k)]^n] + C_m[\rm{rnd}(-1,1)]
\label{eq:channelnoise}
\end{equation}
The function $p(z)$ asymmetrically and non-linearly amplifies $d(t)$, introduces phase distortions, inter-symbol interference, as well as a random noise to generate $u(t)$. We train the network the function $q(z) = p^{-1}(z)$, so that it can recover $d(t)$ from $u(t)$. After training, we test the network and find that for even small size networks ($N=20$), the signal can be extracted with high fidelity from severely distorted signals (Fig. \ref{fig:cheq}b).  In the presented simulation, the Symbol Recovery Rate ($SRR = 1-\displaystyle\frac{|y(t)-d(t)|}{|u(t)-d(t)|}$) was $94.28\%$. From multiple simulations on a wide variety of models for $p(z)$, we have found the $SRR$ to lie in the $90-100\%$ range. More complex filter designs with stacked networks may help increase the performance of such filters. 

This task shows the possibility of building highly compact and energy-efficient dynamically trainable neuro-adaptive filters using ASNs. Such filters can find wide applications in SWaP (size-weight-and-power) constrained environments such as IoT, sensor networks, and self-driving automotives and UAVs.

\section*{Acknowledgment}
This work was partially supported by the NSF I/UCRC on Multi-functional Integrated System Technology (MIST) Center IIP-1439644, IIP-1738752 and IIP-1439680.

\end{document}